\documentclass[11pt,a4paper]{article}
\pdfoutput=1
\usepackage[utf8]{inputenc}
\usepackage{verbatim}
\usepackage{amsthm}
\usepackage{amsmath}
\usepackage{amssymb}
\usepackage{graphicx}
\usepackage{esint}
\usepackage{placeins}
\usepackage{siunitx}
\newcommand{\di}[1]{\text{#1}}

\newcommand{\sunmassesperh}[1]{\SI{#1}{M}_{\odot}\text{/h}}
\usepackage{jcappub}

\begin{document}


\title{The effect of interacting dark energy on local measurements of the Hubble constant}

\author[a]{Io Odderskov,}
\author[b]{Marco Baldi,}
\author[c]{Luca Amendola}

\affiliation[a]{Department of Physics and Astronomy\\
 University of Aarhus, DK-8000 Aarhus C, Denmark}

\affiliation[b]{Dipartimento di Fisica e Astronomia,\\
 Alma Mater Studiorum Universit\`a di Bologna, I-40127 Bologna, Italy}

\affiliation[c]{Institute of Theoretical Physics\\
                University of Heidelberg, 69120 Heidelberg, Germany}
                
\emailAdd{isho07@phys.au.dk}

\date{\today}
\abstract{In the current state of cosmology, where cosmological parameters are being measured to percent accuracy, it is essential to understand all sources of error to high precision. In this paper we present the results of a study of the local variations
in the Hubble constant measured at the distance scale of the Coma Cluster, and test the validity of correcting for the peculiar velocities predicted by gravitational instability theory. The study is based on N-body simulations, and includes models featuring a coupling between dark energy and dark
matter, as well as two $\Lambda$CDM simulations with different values of $\sigma_8$. It is found that the variance in the local flows is significantly larger in the coupled models, which increases the uncertainty in the local measurements of the Hubble constant in these scenarios. By comparing the results from the different simulations, it is found that most of the effect is caused by the higher value of $\sigma_8$ in the coupled cosmologies, though this cannot account for all of the additional variance. Given the discrepancy between different estimates of the Hubble constant in the universe today, cosmological models causing a greater cosmic variance is something that we should be aware of.}
\maketitle

\section{Introduction}

Today, cosmological parameters are being measured with an accuracy
of a few percent. This makes it possible to constrain cosmological
models by comparing independent estimates with each other under the
assumption of some specific model. Recently, this led to questioning
the standard cosmological scenario, when the Planck Collaboration found that measurements
from the CMB under the assumption of the $\Lambda$CDM model led to
a surprisingly low value of the Hubble constant, more than two standard
deviations from the one measured in the local universe \cite{Riess:2011,Planck:2013}.
One possible solution to this that did not challenge the $\Lambda$CDM
model was that the discrepancy was caused by local variations in the
cosmic flow field. This was studied by \cite{Marra:2013,Wojtak:2013,Odderskov2014},
and found to have an insignificant effect at the distance scales probed
by type Ia supernovae. Another possible resolution of the discrepancy,
which has been suggested in \cite{Efstathiou2013}, is that systematics
in the calibration of the distance-luminosity relation for type Ia
supernovae may lead to an overestimate of the Hubble constant. The calibration
used by \cite{Riess:2011} is primarily based on Cepheids, and the
precision of this calibration both limits the accuracy to which the
Hubble constant can be determined and possibly introduces systematics
 leading to an erroneous estimate. 

One of the alternative ways to measure the Hubble constant is to use Cepheids in the Coma Cluster. This would eliminate
a step on the distance ladder and the possible error associated with
the calibration \cite{Gregg2012a,Macri2013}. But at this much smaller
distance, the effect of the local variations is much larger. 

In this paper, we study the local variations associated with a determination
of the Hubble constant using Cepheids in the Coma Cluster. We further
study the model dependence of this method. This is done using a set
of publicly available cosmological simulations and halo catalogues
from the CoDECS project%
\footnote{``Coupled Dark Energy Cosmological Simulations'', the halo catalogues
are available at http://www.marcobaldi.it/CoDECS%
}, which include a coupling between a dark energy scalar field and
dark matter particles \cite{Baldi2010,Baldi2011}. We show
that the uncertainty associated with the local variations is model dependent, and that the presence of
a coupling between dark matter and dark energy would increase the
variance among the locally measured Hubble constants by about
$1\%$. A substantially larger fraction of the observers in cosmologies
with a strong coupling would significantly over- or underestimate
the expansion rate. By performing the same analysis on a standard $\Lambda$CDM simulation featuring a high value of $\sigma_8$, we find that most of the effect can be ascribed to the higher value of $\sigma_8$ arising in the coupled cosmologies as a consequence of the fifth-force mediated by the dark energy field, in order for them to be consistent with constrains from the CMB. However, this does not account for all of the added variance.  

We also test the procedure of subtracting the component of peculiar
velocities predicted by gravitational instability theory, and discuss
the validity of this approach in the coupled cosmologies. We show that the correction reduces the variance in all the considered cosmologies, when the velocities are smoothed over a sufficiently large radius. 

The paper is organized as follows: In section \ref{sec:Coupled-dark-energy-cosmologies},
we provide a short description of coupled dark energy cosmologies.
The simulations and halo catalogues are described in section \ref{sec:Simulations-and-halo-catalogues}.
In section \ref{sec:Method} we describe how the distributions of
local Hubble constants are found by identifying mock observers in
the halo catalogues, who each carry out a mock observation in order
to determine the Hubble constant in their local universe. We also
explain the subsequent correction procedure carried out by each observer.
In section \ref{sec:Results-and-discussion} we present and discuss
our results, and we conclude in section \ref{sec:Conclusions}.

\section{Coupled dark energy cosmologies\label{sec:Coupled-dark-energy-cosmologies}}

\noindent If the dark energy component of the universe is coupled
to some of the matter species, as first proposed in
\cite{Wetterich1995}, this would provide a possibility for gaining
a deeper insight into this mysterious component and to address some of the naturalness problems associated with the onset of the cosmic acceleration. To be able to
detect such a coupling, it is essential to understand what imprint
it would make on observables, for example the density field and the
velocities of its tracers \cite{Amendola2000,Baldi2011a}. The cosmologies considered in this paper
are based on an evolving scalar field, playing the role of the dark
energy, which couples to dark matter.
See \cite{Amendola2000,Amendola2004,Amendola2008,Baldi2010,Baldi2010a,Baldi2011,Baldi2011a,Baldi2011b}
for further details about the motivation and evolution of such cosmologies.
The interaction gives rise to a fifth force, acting between dark matter
particles, which is mediated by the dark energy scalar field. This
modifies the processes of both linear and non-linear structure formation,
and it is therefore natural to consider a possibly important modification
of the local variations in the expansion rates found at different
positions in the universe.

The coupling modifies the evolution equations for dark matter and
dark energy by adding a source term in each of the respective dynamic equations (see
\cite{Baldi2011}):

\begin{alignat}{1}
\dot{\rho_c}+3H\rho_c & =-\sqrt{\frac{16\pi G}{3}}\beta_c(\phi)\rho_{c}\dot{\phi},\nonumber \\
\ddot{\phi}+3H\dot{\phi}+\frac{dV}{d\phi} & =\sqrt{\frac{16\pi G}{3}}\beta_{c}(\phi)\rho_c.
\end{alignat}
Here, $\rho_c$ is the density of cold dark matter, $H$ is the Hubble constant, $G$ is Newtons gravitational constant, $\phi$ is the scalar field, $V(\phi)$ is the potential in which it evolves, and the coupling of the scalar field to cold dark matter is described by the coupling function $\beta_c(\phi)$. Overdots denote derivatives with respect to cosmic time. The models considered in this paper are a subset of the ones included in the CoDECS project, corresponding to a set of flat cosmologies with different assumptions about the potential $V(\phi)$ in which the scalar field evolves, and about the coupling function $\beta_{c}(\phi).$
We use the same naming convention as in the CoDECS project, where
the first part of the name of a model describes the form of $V(\phi)$,
and the last part describes the strength of the coupling function
$\beta_{c}(\phi).$ The models considered are:

\begin{align}
\Lambda\text{CDM:}\quad\quad & V(\phi)=A &  & ; &  & \beta_{c}(\phi)=0,\label{eq:bg:LCDM}\\
\text{EXP00[1,3]:}\quad\quad & V(\phi)=Ae^{-\alpha\phi} &  & ; &  & \beta_{c}(\phi)=\beta_{0},\label{eq:bg:EXP}\\
\text{EXP008e3:\quad\quad} & V(\phi)=Ae^{-\alpha\phi} &  & ; &  & \beta_{c}(\phi)=\beta_{0}e^{\beta_{1}\phi},\label{eq:bg:EXPe}\\
\text{SUGRA003:}\quad\quad & V(\phi)=A\phi^{-\alpha}e^{\phi^{2}/2} &  & ; &  & \beta_{c}(\phi)=\beta_{0},\label{eq:bg:SUGRA}
\end{align}
with the values of $\alpha$, $\beta_0$ and $\beta_1$ given in table \ref{tab:parameters}. See \cite{Amendola2000} for more details about the models with constant
coupling (EXP001 and EXP003), \cite{Baldi2010a} for details about
the model with variable coupling (EXP008e3) and \cite{Baldi2011b}
for details about the SUGRA model.

As shown in \cite{Baldi2010a,Amendola2004}, the relation from linear perturbation theory relating the density parameter for matter, $\Omega_m$, and the growth rate, $f(a)$: $f(a)=\Omega_m^{\gamma}$ with $\gamma=0.55$
\cite{Peebles}, no longer holds in coupled models. In \cite{Baldi2010}
it was found that a good fit can be achieved with the formula

\begin{align}
f(a)\approx\Omega_m^{\gamma}\left(1+\gamma\frac{\Omega_c}{\Omega_{m}}\epsilon_{c}\beta_{c}^{2}\right),\label{eq:modifiedgrowth}
\end{align}
where $\Omega_c$ is the density parameter for cold dark matter, with $\gamma=0.56$ and $\epsilon_{c}=2.4.$ For the models with an exponential potential, EXP001, EXP003 and EXP008e3, the normalization of the scalar field is $\phi(z=0)=0$. This means that $\beta_c(z=0)=\beta_0$ for each of the coupled models.

\section{Simulations and halo catalogues\label{sec:Simulations-and-halo-catalogues}}

The simulations in the CoDECS project have been created using a modified version of GADGET \cite{Springel:2005}.
The implementation of the modifications is described in \cite{Baldi2010}.
The simulations which are used in this study have been run in periodic boxes of
$\SI{1}{Gpc/h}$, containing $1024^{3}$ CDM particles and the same
number of baryon particles, corresponding to particle masses at $z=0$ of $m_c=\sunmassesperh{5.84e10}$
and $m_{b}=\sunmassesperh{1.17e10}$, respectively. The only difference
between the two types of particles is that the baryons do not couple
to the scalar field. The softening length in the simulations is $\epsilon_{s}=\SI{20}{kpc/h}$.

All the simulations share the same set of parameters at the present time, except for the amplitude of density perturbations, $\sigma_{8}$, which has been normalized to the value it had at the time of last scattering ($z_{CMB}\approx1100$), in order to be consistent with constraints from the CMB. Therefore, the different growth rates result in different values of $\sigma_{8}$ today.

Ideally, all of the simulations should be run in different versions with different cosmological parameters, in order to check the importance of each of these on the results. However, this is computationally expensive with simulations of these sizes, and beyond the scope of this work. The most important parameter in this regard is $\sigma_8$, which differs much between the simulations and can be expected to be responsible for a large part of the difference in their velocity fields, which causes the local variations in $H_0$. For the purpose of distinguishing the effect of $\sigma_8$ from separate effects caused by the coupling, we include in the studied models a further realisation of $\Lambda$CDM in which $\sigma_8$ has the same value as in the strongly coupled model, EXP003 (see table \ref{tab:parameters}). This model will be referred to as $\Lambda$CDM\_HS8.  

The initial conditions have been created by displacing particles from a homogeneous glass distribution using the Zel'dovich approximation \cite{Zeldovich}, in order to match a random-phase realization of the linear matter power spectrum (assumed to be Gaussian, with a spectral index of
$n_{s}=0.966$). The simulations were started at an initial redshift of $z_i=99$. The phase of displacements were chosen to be the same for each model, so that the large scale
cosmic structures are the same in all the simulations.

The simulation parameters correspond to the 7th year results of the
Wilkinson Microwave Anisotropy Probe \cite{Komatsu2011}: The universe
is assumed to be flat, and $h=0.703,\Omega_c=0.226,\Omega_{\Lambda}=0.729,\Omega_{b}=0.0451$. This results in particle masses of $m_{c}=\sunmassesperh{5.84e10}$ for the CDM particles and $m_b = \sunmassesperh{1.17e10}$ for the baryons at $z=0$ (the particle masses vary with redshift in the coupled cosmologies).
Only the halo catalogues at $z=0$ have been used. The parameters
are summarized in table \ref{tab:parameters}.

The halo catalogues have been generated using the SUBFIND algorithm \cite{Springel2001}. First, particle groups were identified by running a Friends-of-Friends algorithm over the CDM particles only, with a linking length of $0.2$ times the mean particle distance. Only groups with at least 32 CDM particles have been retained, and baryonic particles having as nearest CDM neighbour a group member where subsequently attached to the group. Next, gravitationally bound substructures (halos)
within the FOF-groups were identified by SUBFIND. In this second step, only halos with at least 20 bound
particles (CDM as well as baryons) were kept. The mass of a group or a halo was calculated as
the sum of the masses of all the particles it contains. 

\begin{table}
\centering{}%
\begin{tabular}{lccccc}
\hline 
Model & A & $\alpha$ & $\beta_{0}$ & $\beta_{1}$ & $\sigma_{8}$\tabularnewline
\hline 
$\Lambda$CDM & - & - & - &  & 0.809\tabularnewline
$\Lambda$CDM\_HS8 & - & - & - &  & 0.967\tabularnewline
EXP001 & 0.0218 & 0.08 & 0.05 & 0 & 0.825\tabularnewline
EXP003 & 0.0218 & 0.08 & 0.15 & 0 & 0.967\tabularnewline
EXP008e3 & 0.0217 & 0.08 & 0.4 & 3 & 0.895\tabularnewline
SUGRA003 & 0.0202 & 2.15 & -0.15 & 0 & 0.806\tabularnewline
\hline 
\end{tabular}
\caption{Cosmological parameters for the simulations. The sidelength of the
box is $1\di{Gpc/h},$ the number of particles is $2\cdot1024^{3},$
and $h=0.703,\Omega_c=0.226,\Omega_{\Lambda}=0.729,\Omega_{b}=0.0451$.
The spectral index for the Gaussian initial conditions is $n_{s}=0.966$,
and the normalization of scalar perturbations is $\mathcal{A}_{s}=2.42\times10^{-9}$.}
\label{tab:parameters}
\end{table}

\section{Method\label{sec:Method}}

Using the halo catalogues of the N-body simulations described above,
we estimate the distribution of Hubble constants that would be observed
by observers who find themselves in a position in the universe similar
to ours. The spread of this distribution contributes to the uncertainty associated with the method.

\subsection*{\noindent Mock observers and observations}

We use the halo catalogues from the CoDECS project, described in the previous section, to mimic the observations
of the Coma Cluster. At first, in each simulation we choose 1000 observers
with a position which has the same characteristics as our position
in the universe. This is done by identifying the halos with a mass
similar to that of the Local Group (in the mass range $\SI{1e12}-\sunmassesperh{1e13}$)
which are subhalos of a group with a mass similar to that of the Virgo
Supercluster (in the mass range $\SI{5e14}-\sunmassesperh{5e15}$).
Each observer makes an observation of a halo at a distance of $65-75\di{Mpc/h}$.
Since $h=0.703$, this corresponds to the distance to the Coma
Cluster ($99\di{Mpc}=70\di{Mpc/h}$). The observer then estimates
the Hubble constant from the measured distance, $r$, and radial velocity,
$v_r$, of the halo as

\begin{equation}
H_{loc}=\frac{v_r}{r}.
\end{equation}
The velocity is the sum of the velocity due to the expansion, given by $H_0r$, and the radial part of the peculiar velocity of the halo. This corresponds to the velocity that would be deduced from the measured redshift.

The peculiar velocities are generally assumed to be generated by the process of structure formation. This can be described using linear perturbation theory, as will be further discussed in the next section. However, this is only valid when the matter distribution is smoothed on sufficiently large scales. The Cepheids at the center of the Coma Cluster are assumed to be at rest with respect to the cluster, and therefore follow its overall motion,\footnote{This assumption might be problematic, as results from large scale surveys indicate that central galaxies might not be at rest at the halo centers. See for example \cite{Guo2015}.} which can be found as the center of mass motion of its constituents. In order to make the observations in the mock catalogues fit the observation of the Coma Cluster as accurately as possible, we measure these velocities in the following way: Each observer starts by identifying the most massive halo in the chosen distance range, and subsequently finds all halos within a radius $R$ of this halo. Then, the center of mass velocity of this group of halos is calculated, which corresponds to smoothing the velocity field on a scale given by $R$. We investigate how the results depend on the scale by using different radii for the calculation of the center of mass motion, ranging from $R=0$, equivalent to no smoothing, to $R=\SI{15}{Mpc/h}$, which is significantly larger than the radius of the Coma Cluster. 

%

The observation carried out by each observer is illustrated in figure
\ref{fig:observation}.

\subsection*{Correcting for local flows}

In \cite{Freedman2001}, the observed velocities are corrected
by subtracting the predicted velocity as calculated from a sum over
the nearby attractors (The Virgo Supercluster, The Great Attractor
and the Shapley Concentration \cite{Mould2000,Freedman2001}).
Thanks to large galaxy surveys and improved computer resources, today
it is possible to integrate the entire density field in the nearby
universe, at least to the degree that it can be accurately probed
by tracers such as galaxies. This approach is for example described
in \cite{Neill2007} using the IRAS PSCz
galaxy redshift survey.

The predictions for the local flows are found by applying linear perturbation
theory to the density field. As well known \cite{Peebles}, the peculiar
velocity at a position ${\bf {r}}$ is related to the gravitational
field through

\begin{align}
{\bf v}({\bf r})=\frac{2}{3}\frac{f(\Omega)}{\Omega_{m}H(z)}{\bf g}({\bf r}),
\end{align}
where $\Omega_{m}$ is the density parameter for matter, $H(z)$
is the Hubble constant at redshift $z$, and $f(\Omega)$ is the growth
rate. The gravitational field ${\bf g}({\bf r})$ can be found by integrating the overdensity
$\delta_m({\bf r})\equiv\frac{\rho_m({\bf r})-\bar{\rho}_m}{\bar{\rho}_m}$, where $\rho_m(\bf r)$ is the matter density at position $\bf r$, and $\bar{\rho}_m$ is its mean value:

\begin{align}
{\bf g}({\bf r})=G\bar{\rho}_m\int\frac{{\bf r}'-{\bf r}}{|{\bf r}'-{\bf r}|^{3}} \delta_m({\bf r}')d^{3}{\bf r}'.
\end{align}
The overdensity field can be related to the halo overdensity
field as $\delta_{h}=b_{h}\delta_m$, where $b_{h}$ is the halo bias.
Using this, and combining the two equations above,
it is found that the peculiar velocity from linear perturbation theory
at position ${\bf r}$ is given by

\begin{align}
{\bf v}({\bf r})=\frac{H(z)f(\Omega)}{4\pi\bar{\rho}_{m}b_{h}}\int\frac{{\bf r}'-{\bf r}}{|{\bf r}'-{\bf r}|^{3}}{\bf \delta}_{h}({\bf r}')d^{3}{\bf r}',
\end{align}
where we have also used the Friedmann
equation, $H^2(z)=\frac{8\pi G}{3}\frac{\rho_{m}}{\Omega_{m}}$. It is assumed that the monopole term of the integral does not
contribute, as will be the case if the integral is over all of space
or over a sphere around the halo at position ${\bf r}$. The density field
of the halos in the simulation can be represented as a sum of delta
functions at the halo positions, and thereby the expression above
turns into 

\begin{align}
{\bf v}({\bf r})=\frac{H(z)f(\Omega)}{4\pi\bar{\rho}_{m}b_{h}}\sum_{i}\frac{{\bf r}_{i}-{\bf r}}{|{\bf r}_{i}-{\bf r}|^{3}}m_{i},\label{eq:vg}
\end{align}
where $m_{i}$ is the mass of the $i$th halo. In our analysis, we
assume $b_{h}=$1. We use the growth rates given in equation \ref{eq:modifiedgrowth}.

\begin{figure}
\includegraphics[width=1\textwidth]{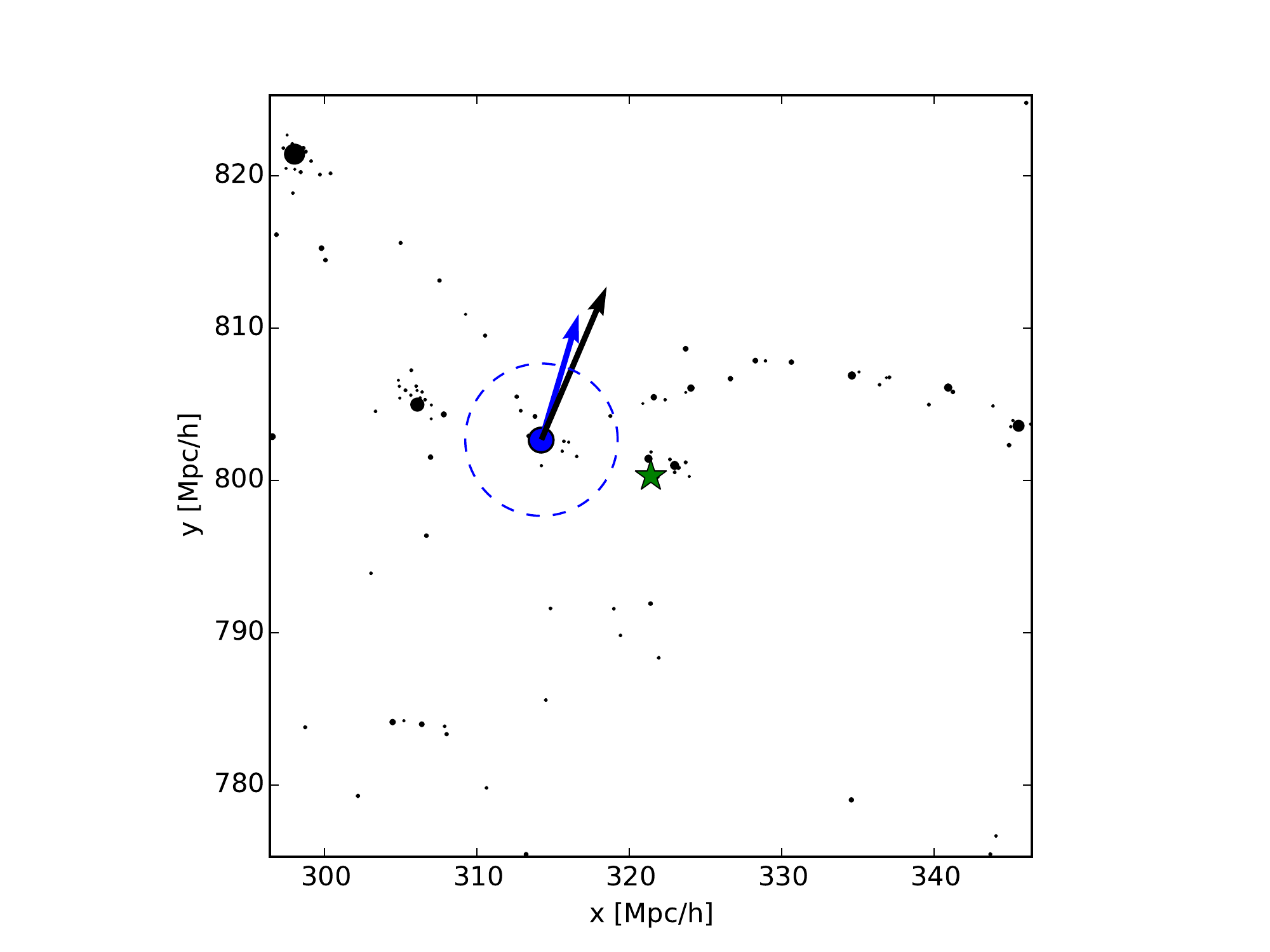}\caption{An illustration of the observation performed by each observer. The
observer is marked with a green star. It chooses the most massive
halo in the distance range $65-\SI{75}{Mpc/h}$. The peculiar component of the center of mass motion of
all halos within $R=\SI{5}{Mpc/h}$ (in the case displayed above)
is then calculated (blue arrow), and the radial part is used for the
estimation of the Hubble constant. The black arrow is the velocity
calculated with equation \ref{eq:vg}. The slice is $\SI{10}{Mpc/h}$
thick, and is centered on the observer.\label{fig:observation}}
\end{figure}

Linear perturbation theory only gives an accurate description of the velocity field when it is smoothed with a sufficiently large smoothing radius. When the smoothing scale $R$ is small, a massive halo just outside the observed sphere (the blue circle in figure \ref{fig:observation}) will lead to extreme predictions for the velocity, due to the distance in the denominator of equation \ref{eq:vg}. Such velocities will be only weakly correlated with the actual velocities, because they are determined by processes that are not captured by linear perturbation theory, such as virialization. In these cases, the correction procedure will lead to estimates of the Hubble constant very far from the true value. Therefore, the correction procedure can be expected to give very bad results in the case of a small smoothing scale.

Due to the intimate relation  between the density field and the peculiar velocity field, the amount of clustering, measured by $\sigma_8$, can be expected to significantly affect the results. This issue is also discussed in \cite{Marulli2011}, where the effect of interacting dark energy on redshift space distortions is studied, and found to be strongly degenerate with the value of $\sigma_8$ at scales larger than $5-\SI{10}{Mpc/h}$. Since the scale considered in this study is much larger than this ($\sim \SI{70}{Mpc/h}$), we expect our results to be strongly dependent on the values of $\sigma_8$ featured in the different models.

\subsection*{\noindent Distribution of local Hubble constants}

\noindent 
The difference in the measured Hubble constants at different locations in the universe is a consequence of the inhomogeneities in the matter distribution, which induces peculiar velocities. The matter distribution, described in terms of the overdensity parameter, $\delta_m=\frac{\rho_m(\bf r)-\bar{\rho}_m}{\bar{\rho}_m}$, is necessarily characterized by a skewed distribution, since the allowed values of $\delta_m$ ranges from $-1$ to $\infty$. In agreement with this, it has been found that it is well described by a log-normal distribution \cite{Coles1991}. As in \cite{Marra:2013}, we assume that local Hubble constants follow the same distribution as the inhomogeneities in the matter distribution, since this is the source of the variations in the Hubble constants. Therefore, we fit the distribution of local Hubble constants in the simulations by a log-normal distribution:

\begin{align}
f(x)=\frac{1}{x\sigma\sqrt{2\pi}}\exp{\Biggl[\frac{-(\ln{x}-\mu)^{2}}{2\sigma^{2}}\Biggr]},
\end{align}
where in our case $x=H_{loc}/H_{0}$. The mean and spread of $\ln(x)$
are given by the fitting parameters $\mu$ and $\sigma$. In table
\ref{tab:distributions}, we show the corresponding mean and spread of $x,$ namely $\mu_{x}=e^{\mu}$ and $\sigma_{x}=\mu_{x}\cdot\sigma$,
where the last part follows from the law of propagation of errors.

\section{Results and discussion\label{sec:Results-and-discussion}}

\begin{figure}
\includegraphics[width=\textwidth]{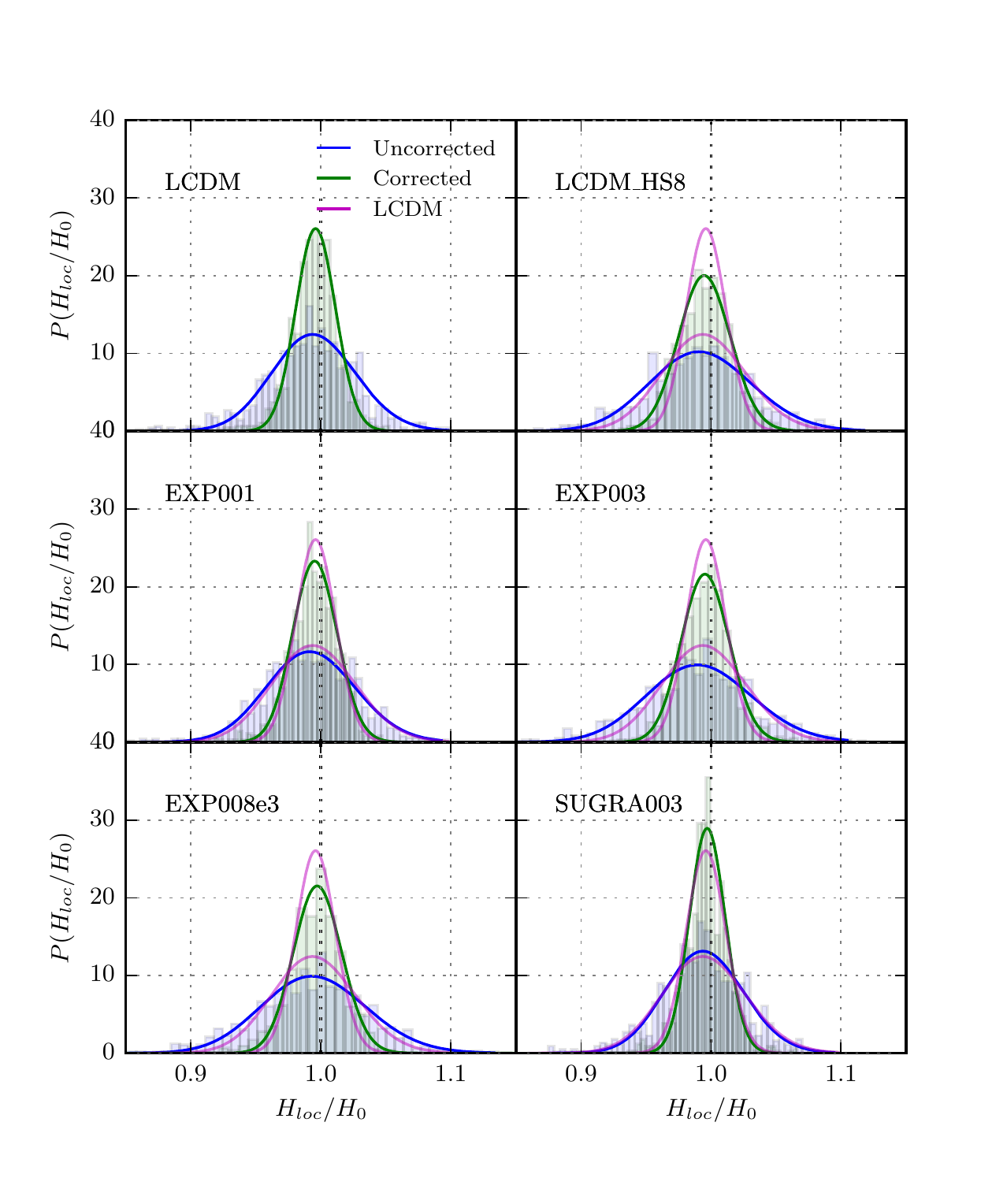}
\caption{The distribution of the measured Hubble constants among the halos with
characteristics similar to that of the local group in each of the considered models. The CoM scale is $R=\SI{5}{Mpc/h}$ in all cases. The blue histograms corresponds to the uncorrected distributions, the green histograms to the corrected distributions. The lines are the fitted log-normal distributions. The fitted distributions for the $\Lambda$CDM model are shown in magenta for comparison.  
\label{fig:Hdists_all}}
\end{figure}

In figure \ref{fig:Hdists_all}, we show the distributions of the Hubble constants measured by the different observers in each of the considered models. For each model, we show both the original (blue) and the corrected (green) distribution, which is obtained by subtracting the peculiar velocity predicted by linear theory, equation \ref{eq:vg}, before calculating the Hubble constant. Each model is compared to the distributions in the standard $\Lambda$CDM model, which are indicated in magenta. We have chosen a smoothing radius $R=\SI{5}{Mpc/h}$, since this approximately corresponds to the size of the Coma Cluster. 

\begin{figure}[htb!]
\center\includegraphics[width=\textwidth]{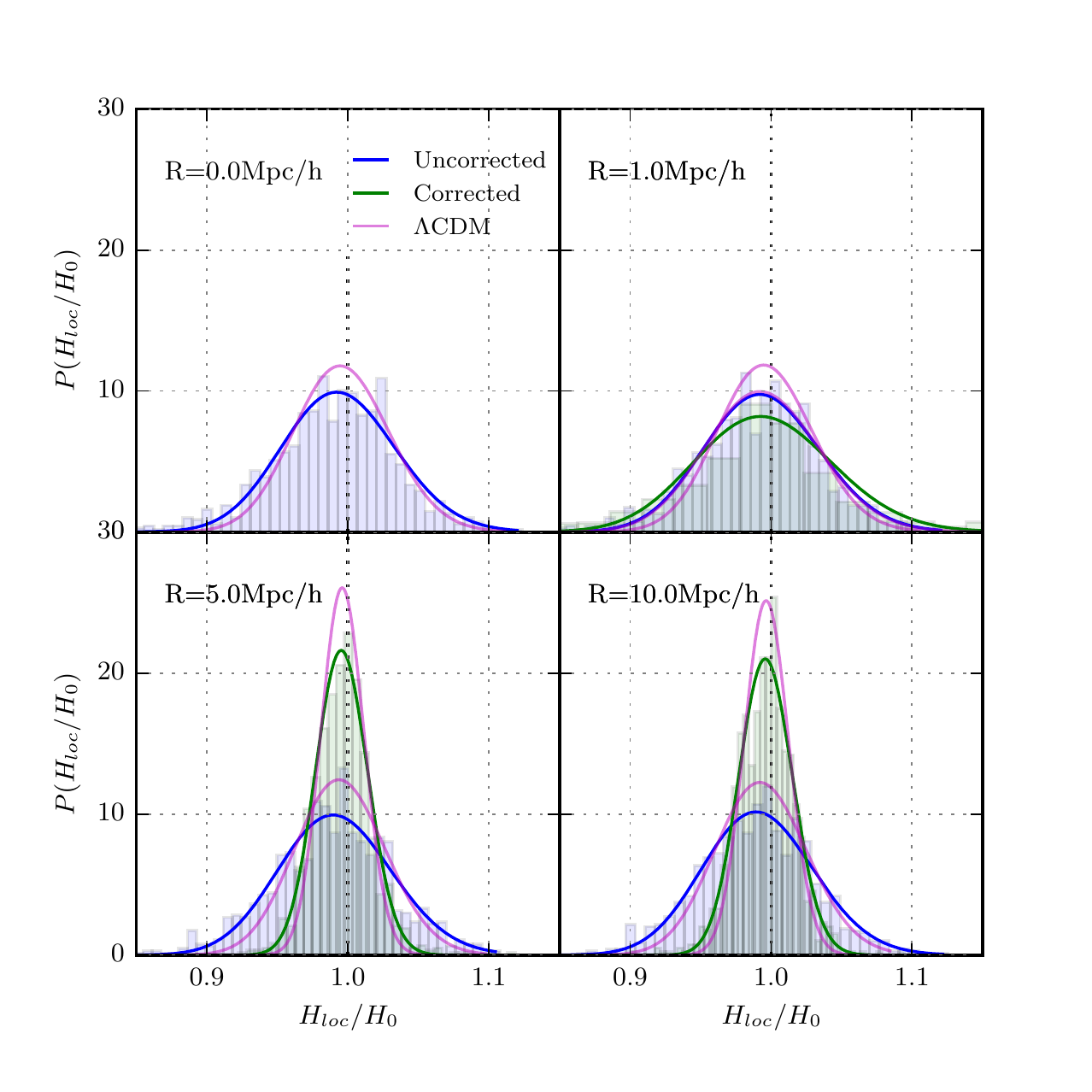} \caption{The distribution of the measured Hubble constants among the halos with
characteristics similar to that of the local group in the strongly
coupled model (EXP003). The blue histograms are the distributions before the correction procedure is applied, the green histograms corresponds to the corrected distributions (not shown for the case without any smoothing, $R=0$). The lines corresponds to the fitted log-normal distributions. The fitted distributions for the $\Lambda$CDM model are shown in magenta for comparison.
\label{fig:Hdists_EXP003} }
\end{figure}

The dependence on the smoothing length is illustrated in figure \ref{fig:Hdists_EXP003}, where the corrected and uncorrected distributions are shown for smoothing radii of $0$, $\SI{1}{Mpc/h}$, $\SI{5}{Mpc/h}$, and $\SI{10}{Mpc/h}$, respectively, for the EXP003 model. 



\begin{figure}
\includegraphics[width=\textwidth]{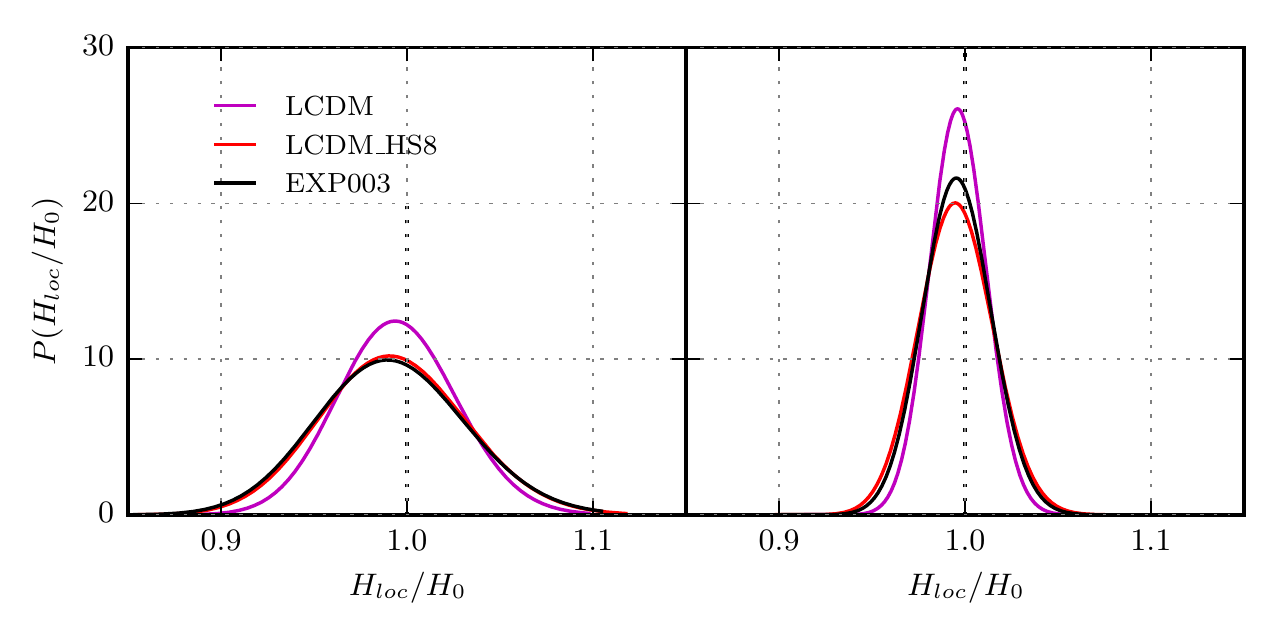}
\caption{Comparison between fitted distributions for uncorrected (left), and corrected (right) distributions of the locally measured Hubble constants. It is seen that most of the difference between the distributions found in $\Lambda$CDM and in EXP003 can be attributed to the higher value of $\sigma_8$ in the coupled model.  
\label{fig:Hdists_sigma8}}
\end{figure}

The parameters for the fitted distributions are given in table \ref{tab:distributions}.
Before correcting for the linear flows, the models with a strong coupling
(EXP003 and EXP008e3) show a spread in the locally measured Hubble
constants between 0.6\% and 0.8\% larger than the spread found in the $\Lambda$CDM
model. It is perhaps even more interesting to consider the tails of the histograms. A substantial fraction, 13.4\%, of the observers in the $\Lambda$CDM model would over- or underestimate the Hubble constant by more than 5\%, and this number increases to respectively 24.4\% and 22.9\% in the cosmologies with a strong coupling.
The distributions in the weakly coupled cosmologies are very close
to the one found for $\Lambda$CDM.

For the case of the $\Lambda$CDM\_HS8 model, featuring a high value of $\sigma_8$, the distributions are very similar to those found for EXP003, which has the same value of $\sigma_8$. This is shown in figure \ref{fig:Hdists_sigma8}, in which the fitted distributions for EXP003, $\Lambda$CDM, and $\Lambda$CDM\_HS8 are compared. This indicates that most of the difference between the coupled and uncoupled models can be attributed to the different values of $\sigma_8$. There is, however, a small residual difference. This is most notable when considering the tails of the histograms: The fraction of observers who over- or underestimate the Hubble constant by more than 5\% is a few percent higher in the EXP003 model than in the $\Lambda$CDM model with the same value of $\sigma_8$. Furthermore, the model with a variable coupling, EXP008e3, also has a wider distribution than $\Lambda$CDM\_HS8, despite having a lower value of $\sigma_8$. This is similar to the results presented in \cite{Lee2011}, in which the pairwise infall velocity of clusters is found to be greatest in the simulation which features a variable coupling. In \cite{Lee2011}, this result is further investigated by considering the first order equations from \cite{Amendola2004} for the density and velocity perturbations in coupled dark energy models. By solving these, it is shown that the biggest perturbations in the peculiar velocity field at $z=0$ are found in the exponentially coupled model, even though the corresponding enhancement of the density perturbations (as measured by $\sigma_8$) is significantly smaller than in the EXP003 model. Hence, this result is in agreement with what one should expect from linear perturbation theory.\\
\newline\noindent
When varying the radius used for the CoM calculation, we see that increasing the radius generally leads to better results for the correction procudure. This is expected, as velocities
on larger scales will be better approximated by linear perturbation theory. However,
when the CoM-scale is increased to $15\di{Mpc/h}$, the correction procedure becomes less
effective. We attribute this to the effect of the finite box size. It is seen that the correction slightly underestimates the peculiar
velocity caused by the presence of the halo in which the observer
is situated, by the fact that also the corrected distribution has
a mean value less than one. This might be caused by the choice to
set the halo bias $b_{h}$ equal to one. Apart from this, the correction
procedure efficiently reduces the spread of the distribution for both
the standard and the coupled cosmologies. It is also seen that linear
theory only reduces the uncertainty at scales above a couple of Mpc.
However, structures of a size similar to that of the Coma Cluster
(with a radius a little above $5\di{Mpc/h}$) have a peculiar motion which
appears to be well described by this theory. 

\begin{center}
\begin{table}
\center %
\begin{tabular}{lllllll}
Model & $\sigma_8$ & R  & $\mu_{x}$  & $\sigma_{x}$  & $P(\frac{H_{loc}}{H_{0}}<0.95)$  & $P(\frac{H_{loc}}{H_{0}}>1.05)$ \tabularnewline
\hline 
& & 0.0 & 99.6 (-) & 3.4 (-) & 10.8 (-) & 5.3 (-)\tabularnewline
& & 1.0 & 99.6 (99.4) & 3.4 (4.0) & 10.4 (17.3) & 4.7 (7.7)\tabularnewline
\textbf{$\Lambda$CDM} & 0.809 & 5.0 & 99.5 (99.6) & 3.2 (1.5) & 10.3 (1.6) & 3.1 (0.6)\tabularnewline
& & 10.0 & 99.3 (99.7) & 3.3 (1.6) & 9.9\phantom{0} (1.2) & 4.7 (0.2)\tabularnewline
& & 15.0 & 99.2 (99.7) & 3.3 (1.8) & 11.3 (1.5) & 4.4 (0.6)\tabularnewline
 &  &  &  & \tabularnewline
& & 0.0  & 99.3 (-) & 4.0 (-) & 17.7 (-) & 7.3 (-)\tabularnewline
& & 1.0  & 99.4 (99.5) & 4.1 (4.9) & 18.0 (23.6) & 8.0 (16.2)\tabularnewline
\textbf{EXP003} & 0.967 & 5.0  & 99.1 (99.6) & 4.0 (1.8) & 17.0 (1.5) & 7.4 (1.4)\tabularnewline
& & 10  & 99.1 (99.6) & 3.9 (1.9) & 18.0 (1.5) & 5.8 (0.4)\tabularnewline
& & 15  & 99.2 (99.7) & 4.0 (2.1) & 14.7 (3.5) & 6.0 (1.4)\tabularnewline
 &  &  &  & \tabularnewline
\textbf{$\Lambda$CDM\_HS8} & 0.967 & 5.0 & 99.2 (99.6) & 3.9 (2.0) & 15.3 (2.3) & 6.3 (0.7)  \tabularnewline
\textbf{EXP001} & 0.825 & 5.0 & 99.3 (99.6) & 3.4 (1.7) & 10.1 (1.8) & 3.1 (0.4) \tabularnewline
\textbf{EXP008e3} & 0.895 & 5.0 & 99.5 (99.8) & 4.0 (1.9) & 15.8 (3.0) & 7.1 (0.9) \tabularnewline

\textbf{SUGRA003} & 0.806 & 5.0 & 99.5 (99.7) & 3.0 (1.4) & 9.8\phantom{0} (1.3) & 3.8 (0.4)\tabularnewline
\hline
\end{tabular}\caption{Parameters for the fitted distributions (mean and spread of $x=H_{loc}/H_{0}$),
and the probabilities of measuring a Hubble constant
which deviates from the true value by 5\% or more, as calculated directly from the histograms. The numbers in
the parentheses are the parameters after correcting for the linear
velocities induced by the matter distribution. These are omitted for
the case of no smoothing, $R=0$, as non-linear velocities completely dominates
in this case. All the numbers are in percent, except for the CoM-scale, R, which is in $\di{Mpc/h},$ and $\sigma_8$.
\label{tab:distributions}}
\end{table}
\par\end{center}
%
%
%

Future studies might include a more thorough exploration of the peculiar velocities in coupled cosmologies. This subject is also treated in \cite{Fontanot2015}, where differences in the pairwise velocity distributions of the coupled cosmologies are explored. It is briefly discussed how the effects of the coupling compares to the effects caused by the existence of massive neutrinos. The signature from massive neutrinos points in the opposite direction of the change caused by the coupling \cite{Vacca2009}. Therefore, if massive neutrinos were present, this could allow for a large coupling. And since the coupling causes a larger variance, this could potentially explain a discrepancy between the Hubble constant estimated from the CMB and the value measured in the local universe. In the most strongly coupled model considered here, EXP003, 0.8\% of the observers would measure a local Hubble constant more than 8.8\% higher than the true value, corresponding to the discrepancy between the local and the CMB value. This is 4 times as many as in the $\Lambda$CDM case (0.2\%). However, the corresponding fraction in the $\Lambda$CDM\_HS8 model is also 0.8\%, which confirms that the higher value of $\sigma_8$ is responsible for most of the additional variance. \FloatBarrier

\section{Conclusions\label{sec:Conclusions}}

We have investigated local variations in the Hubble constant as measured from Cepheids in the Coma Cluster, using a set of simulations featuring a coupling between a dark energy scalar field and dark matter, as well as two $\Lambda$CDM simulations with different values of $\sigma_8$. In the standard $\Lambda$CDM model, a substantial fraction of the observers (more than 13\%) would over- or underestimate the Hubble constant by more than 5\%. The variance is found to be much larger in the strongly coupled models, such that almost twice as many observers measures a Hubble constant which deviates substantially from the true value. Most of the added variance was found to be attributed to the higher value of $\sigma_8$ in the coupled models. However, especially in the model featuring a variable coupling, there is a residual effect that cannot be explained by $\sigma_8$.

We have tested how the distributions of local Hubble constants are affected by correcting
for peculiar velocities induced by the matter distribution, as predicted
by linear theory of structure formation, also taking into account the modified relation between density and velocity arising in interacting Dark Energy cosmologies. The correction procedure effectively reduces the spread in the distributions in both standard and coupled cosmologies. And although the fraction of observers who would significantly over- or underestimate the Hubble constant is still much larger in the coupled cosmologies after the correction procedure, it is less than 4\% in all the considered cosmologies for a smoothing radius of $\SI{5}{Mpc/h}$.

\section{Acknowledgements}

We acknowledge computing resources from Center for Scientific Computing
Aarhus. L.A. acknowledges support from DFG through the project TRR33 'The Dark Universe'. MB acknowledges support from the Italian Ministry for Education, University and Research (MIUR) through the SIR individual grant SIMCODE, project number RBSI14P4IH.

\bibliographystyle{unsrt}
\bibliography{Cepheids}

\end{document}